\documentclass[twocolumn,aps,prl,showpacs,amsmath,amssymb,psfrac,superscriptaddress]{revtex4}

\usepackage{epsfig}
\usepackage{amsfonts,amsmath,amssymb}\usepackage{graphicx}\begin{document}

\title{DYNAMICAL LOCAL LATTICE INSTABILITIY TRIGGERED HIGH $T_c$ SUPERCONDUCTIVITY.}

\author{Julius RANNINGER}

\affiliation{Institut N\'eel, CNRS and Universit\'e Joseph Fourier, \\
 BP 166, 38042 Grenoble cedex 9, France
 \footnote{julius.ranninger@neel.cnrs.fr}}
 
\date{\today}

\begin{abstract}

High $T_c$ cuprate superconductors are characterized by two robust features: their strong electronic correlations and their  intrinsic dynamical local lattice instabilities. Focusing on exclusively that latter, we picture  their parent state in form of a quantum vacuum representing  an electronic magma in which bound diamagnetic spin-singlet pairs pop in and out of existence in a Fermi sea of itinerant electrons. 
The mechanism behind that resides in the structural incompatibility of two stereo-chemical configurations Cu$^{\rm II}$O$_4$ and Cu$^{\rm III}$O$_4$ which compose the CuO$_2$ planes. It leads to  spontaneously fluctuating Cu - O - Cu valence bonds which establish a local Feshbach resonance exchange coupling between bound and unbound electron pairs. The coupling, being the only
free parameter in this scenario, the hole doping of the parent state is monitored by varying the total number of unpaired and paired electrons, in chemical equilibrium with each other. Upon lowering the temperature to below a certain $T^*$,  bound  and unbound electron pairs lock together in a  local quantum superposition, generating transient localized bound electron pairs and  a concomitant  opening of a pseudo-gap in the single-particle density of states. At low temperature, this pseudo-gap state transits via a first order hole doping induced phase transition into a superconducting state in which the localized transient bound electron pairs get spatially phase correlated. The mechanism driving  that transition is a  phase separation between two phases having  different relative densities of bound and unbound electron pairs, which is reminiscent of the physics of $^4$He - $^3$He mixtures.
\end{abstract} 

\maketitle

\section{I Introduction}
Quite independent on any microscopic mechanism leading to superconductivity, this phenomenon
is  generated by establishing  a macroscopic coherent quantum state in which an 
ensemble of transient bosonic charge carriers (composed of diamagnetic electron-pairs), 
having arbitrary phases in the parent state above $T_c$, undergoes a global spontaneous 
symmetry breaking (SSB). The arbitrary phases of these virtual bosonic entities are 
thereby locked together into a unique global (though arbitrary) phase, the excitations of 
which are symmetry restoring collective Goldstone modes. 
In a current carrying state their existence assures the  persistence of the  
resistance-less conduction through the Anderson-Higgs mechanism, by which they contribute to set
up  a longitudinal  component of the electromagnetic vector potential driving this current, as 
recently reviewed \cite{Ranninger-2012} in  commemorating the centennial  anniversary
of the discovery of superconductivity \cite{Kamerlingh-Onnes-1911}.  The value of the 
critical temperature $T_c$ at which a super-flow sets in, depends however sensibly on how 
this SSB comes about in (i) forming   finite amplitudes of individual bosonic 
entities and (ii) establishing the phase coherence between them in order to construct a
macroscopic coherent quantum state. There are two  ways for that to happen.  

(I) {\it When the strength of the inter-pair phase correlations, locking together the bosonic 
entities is large  compared to the pairing energy}. This is the case for BCS superconductors.
The interaction between the electrons, monitored by the exchange of a phonon, is too weak to 
guarantee real space pairing. Yet, the ensemble of such virtual pairs, existing in form of transient 
Cooper pairs in momentum space, situated in a thin layer around the Fermi surface and 
having arbitrary phases,  can be phase-locked into a macroscopic coherent quantum state through 
a collective process \cite{BCS-1957}. It provides the 
required strength for pairing, mediated by inter-pair phase correlations, engaging 
simultaneously a 
macroscopic number of transient Cooper pairs.  Its resulting $T_c$ is controlled by 
the zero temperature pairing amplitude $\Delta(0)$, tantamount to the energy of the single-particle 
gap $\Delta(0) \simeq 1.76 k_B T^{BCS}_c$  with $T_c$ being given by
\begin{eqnarray}
\label{eq:TcBCS}
T_c^{BCS} &\simeq& \omega_D exp\left(-{1\over \lambda/(1+\lambda)- \mu^*}\right) \\
\label{eq:mu*}
\mu^* &=& \mu / [1+ \mu \, ln(\varepsilon_F/\omega_D)] \\
\label{eq:mu-lambda}
\mu-\lambda &=& \rho(\varepsilon_F) \langle V_{el-ph}(q, \omega=0) \rangle_{FS} \\
\label{eq:Vel-ph}
V_{el-ph}(q, \omega=0)&=&{4\pi e^2 \over q^2 \varepsilon(q, \omega=0)},
\end{eqnarray}
$T_c$ sensibly depends on  the  difference between the attractive phonon-mediated electron-electron
interaction $\lambda$ and the repulsive bare Coulomb interaction $\mu$, given by the 
electron lattice coupling $V_{el-ph}(q, \omega=0)$. 
Appearing in form of the average over the Fermi surface,  $V_{el-ph}$ contributes 
predominantly through its small $q$-components of the static dielectric function 
$\varepsilon(q, \omega=0)$. For q=0 having to be positive, in order to assure global 
crystalline stability, this renders $\mu-\lambda$ repulsive rather than attractive. Pairing 
in the BCS scenario finally occurs  because, generating the superconducting state in such
a collective process, it is the screened Coulomb interaction $\mu^*$ rather than the
bare one which controls it \cite{Morel-Anderson-1962} and for which $\mu^*- \lambda \leq 0$. 
But $\varepsilon(0,0)$  having to be positive  still puts a stringent condition on obtaining 
sizeable values of $T_c^{BCS}$. Setting optimally $\mu = \lambda$ and varying the  Debye frequency 
$\omega_D$ in  Eq. \ref{eq:TcBCS} in order to optimize the value of $T_c^{BCS}$, one obtains  
a maximal $T_c^{BCS}(max) = \varepsilon_F e^{-(4 +3/\lambda)}$, which
for typical values of $\lambda$ and $\varepsilon_F$ can barely exceed 30 K. 

Pines and Nozieres \cite{Pines-Nozieres-1966} pointed out that the causality of the response
of the system's internal total charge to an external test charge, given by the  dielectric 
function $\varepsilon({\bf q}, \omega)$, requires that it is the inverse of it rather than 
the dielectric function itself which has to obey the Kramers-Kronik relation. From that, 
Kirshnitz \cite{Kirshnitz-1976} concluded that, together with 
$Im \, \varepsilon({\bf q},\omega=0)  \leq 0$, a negative finite momentum $\varepsilon({\bf q}, 0)$
not only is not incompatible with overall crystalline stability, but can in fact   
 over-screen  the repulsive  Coulomb interaction by dynamical structural instabilities triggering  diamagnetic pairing correlations on a loca scale \cite{Ginzburg-Kirshnitz-1977,Dolgov-Kirshnitz-Maksimov-1981}. With this insight the systematic search for materials  with incipient 
crystalline instabilities became a priority \cite{Matthias-1971,Vandenberg-1977}. 
${\rm A}15$ compounds, showing displacive Martensitic lattice 
instabilities and attaining a $T_c=25$ K in  Nb$_3$Ge \cite{Gavaler-1973}  re-enforced
this strategy.
 
Similar reasoning,  in the early 1980ties, led our group in Grenoble to investigate 
transition metal oxides, such as Ti$_4$O$_7$ \cite{Lakkis-1976}, which showed that their 
Fermi sea of bare itinerant electrons was unstable towards a charge disproportionated  charge 
density wave (CDW) composed of alternating Ti$^{\rm III}$-Ti$^{\rm III}$  and Ti$^{\rm IV}$-Ti$^{\rm IV}$ diatomic molecular complexes housed inside deformable octahedral ligand environments. 
Given that the 
localized  diamagnetic spin-singlet pairs on such molecular sites form small Bipolarons, we 
proposed that upon doping Ti$_4$O$_7$ with V or Sc the insulating CDW state could be destabilized
and  make the system transit into a  Bipolaronic Superconductor \cite{Alexandrov-Ranninger-1981}, 
assuming  that such \`a priori localized Bipolarons could  be rendered itinerant
and result in a super-fluid phase with a $T_c$ determined by their mass density.

(II)) {\it When the strength of the inter-pair phase correlations is small compared to the pairing 
energy.} In that case bound real-space pairs are formed without having to 
invoke their  condensation, such as in potential Bipolaronic Superconductors. 
In order for those  bound electron pairs to condense into 
a super-fluid state, analogous to that of super-fluid $^4$He II with a $T_c$ being controlled 
by the mass density of super-fluid charge carriers, these bosonic entities have: (i) to be
locally well defined individual particles, not or only weakly overlapping with each other and 
(ii) to  exist in form of itinerant states. 
On a very  general level, the scheme  of real space pair  superconductivity had been addressed 
on the basis of the negative U Hubbard model by numerous authors as was reviewed 
\cite{Micnas-Ranninger-Robaszkiewicz-1990} in the early years of the High 
Temperature Superconductivity era, when it was used to account for the crossover from a 
BCS state to a  Bose-Einstein Condensate (BEC) in the so-called phase 
fluctuation scenario \cite{Emery-Kivelson-1995}.  Given that in 
real materials the pairing of electrons is generally generated by  strong local lattice 
deformations which trap electrons into small localized Bipolarons, it became evident that such  local pairs could not exist  in form of itinerant charge carriers 
\cite{Chakraverty-Ranninger-Feinberg-1998}. Searching a way out of this dilemma led me to
propose that  resonating localized Bipolarons could achieve 
a phase fluctuation driven superconducting state on the basis of their transient nature. It 
is that which  permits them through their  amplitude  fluctuations to  lock 
together their  respective phases in a macroscopic coherent  quantum state. 
The intrinsic metastability  of the 
of the cuprate HTSCs, discovered in 1986 \cite{Bednorz-Mueller-1986}, provides us with  
this  prerequisite encountering transient localized bound electron-pairs well above $T_c$.

Physical realizations of  BEC driven superconductivity had been known   
for some time to occur 
in diamagnetic insulating, respectively semiconducting, parent compounds upon 
substitutionally doping them with cations, or rendering them sub-stoichiometric. 
Their cation-ligand configurations are capable to  sustain superconducting diamagnetic pairing 
fluctuations in very dilute concentrations of charge carriers, typically around 10$^{20}$ per 
cm$^3$ in compounds such as SrTiO$_{3-x}$ with a $T_c \simeq 0.3$ K \cite{Schooly-1965}. Substitutionally doped SrTiO$_3$ i.e. SrTi$_{0.97}$Zr$_{0.03}$O$_3$, 
\cite{Tainsh-Andrikidis-1986} presents a  superconducting BEC with  $T_c \simeq 0.07$ K for a 
charge 
carrier concentration as low as of $4 \times 10^{15}$ per cm$^3$. David Eagles \cite{Eagles-1969}
pointed out that this result is compatible with the classical BCS pair exchange mechanism 
in the limit of very low carrier concentrations. Tony Leggett \cite{Leggett-1980} a few years 
later, in a quest to describe super-fluidity of $^3$He, showed how the  BCS ground state
wave-function in  the weak pairing regime  describes  in the limit of very low carrier
concentrations a BEC of real space pairs. Achieving in that scenario higher concentrations 
with well defined local bosonic bound electron pairs of not too heavy  masses and thus high 
$T_c$'s however had been hampered for  many years.

\section{II Resonating Bipolarons}
Cuprate HTSCs manage to evade the problem related to the mobility of real space pairs 
in crystalline materials and a $T_c$ controlled by  their  mass density. Their non-Fermi
 liquid electronic magma parent state at high temperatures can be monitored over a large 
 regime of carrier concentrations  upon changing, in the chemical synthesis, the 
 relative composition of the basic ingredients: the stereo-chemical Cu$^{\rm II}$O$_4$ and
 Cu$^{\rm III}$O$_4$ molecular clusters. 

There are two energy scales which control the inter-dependence of (i) the onset of the 
pseudo-gap state out of a high temperature electronic magma parent state and  (ii) 
the superconducting state, evolving out of this pseudo-gap state at low temperatures 
through a hole-doping induced phase first order phase transition. 

The first one is given by  the pairing energy, characterized by the temperature $T^*$  signalling 
the opening of pseudo-gap in the single-particle density of states. It is  related to 
the generation of  transient localized diamagnetic spin-singlet pairs on competing with 
each other Cu$^{\rm II}$ - O - Cu$^{\rm II}$  and Cu$^{\rm III}$ - O - Cu$^{\rm III}$ 
valence bonds, generating a finite fluctuating pairing amplitude, which monotonously decreases 
with increasing hole doping.

The second one is given by the energy, required to phase lock together these transient 
localized  electron pairs on adjacent fluctuating molecular clusters. This is achieved 
by their  faculty  to spontaneously decay into a pair of itinerant electrons, which 
establishes this phase correlation through Andreev type scattering processes. 

The strogly hole doping dependent competition between  inter-pair and intra-pair phase 
correlations dictates the phase transition by which the  superconducting state, composed of spatially phase correlated transient localized bound electron-pairs, transits into the insulating pseudo-gap state with phase uncorrelated such bound electron-pairs.

When  the pairing energy outweighs the energy related to the phase stiffness, which  links 
adjacent transient pairs as is the case for low  doped cuprate HTSCs up to the optimally 
doped ones, the decreasing with hole doping $x$ pairing energy $k_BT^*(x)$  weakens the 
local intra-pair phase rigidity and thereby permits to  strengthen the inter-pair phase
stiffness. As a consequence,  $T_c(x)$ increases with 
increasing $x$. Being controlled exclusively by phase fluctuations, $T_c$ scales with 
the mass density of the transient bound electron pairs, as experimentally  established 
in Uemura's universal plot of $T_c$ versus $n_s/m_s$ \cite{Uemura-1989}, obtained from 
positron annihilation studies. Upon approaching the optimal doping  $x_{opt}$, a maximal 
value of $T_c(x_{opt})$
is reached when  amplitude and phase fluctuations corroborate optimally to construct a 
state in which both of those fluctuations are simultaneously minimized.  Beyond $x_{opt}$, the pairing energy $k_BT^*(x)$, decreasing with increasing hole doping, takes over the control of 
a BCS like superconducting state triggered by amplitude fluctuations.

These features are to a certain extent compatible with both of the two robust characteristics
of the cuprate HTSCs: (i) their intrinsic dynamical local instabilities of the crystalline lattice 
and (ii) their strong electronic correlations which characterise a hole doped 
Mott insulator, resulting in Phil Anderson's resonating valence bond (RVB) 
scenario \cite{Anderson-1987}.  Both of these scenarios  lead one to a picture in which 
bare itinerant electrons, moving in planar CuO$_2$ structures, get momentarily  bound in 
form of resonating diamagnetic  spin-singlet pairs on plaquettes composed of four Cu
cations. Under certain provisos,  the RVB scenario can be  mapped  into the phenomenological
effective Boson-Fermion Model (BFM) \cite{Altman-Auerbach-2002} which had been advocated prior
to the discovery of the cuprate HTSCs to capture the physics of metastability 
driven superconductivity. In an early Mean Field analysis \cite{Ranninger-Robaszkiewicz-1985}
of the BFM it indicated for the first time the potentiality of a pseudo-gap state controlled
by amplitude fluctuations in a system  of intrinsically localized transient  bipolarons. The
full implications of that had however been recognized \cite{Ranninger-Micnas-Robaszkiewicz-1989} only once the cuprates superconductors had been  discovered.

Our investigations during the past two decades of meta-stability 
driven pairing in the cuprates on the basis of the BFM permitted us to predict the salient features of this pseudo-gap state: (1) the anomalous temperature dependence of this pseudo-gap
\cite{Ranninger-Robin-Eschrig-1995}, (2) the remnant Bogoliubov modes 
\cite{Domanski-Ranninger-2003} and (3) to account for its   
transient Meissner effect \cite{Devillard-Ranninger-2000}). With this background we restrict
ourselves now to focus on the scenario of lattice metastability triggered superconductivity,
which  requires, to start with, to  conjecture a corresponding to it 
quantum vacuum parent state. The  physics of the cuprate HTSCs, manifest in their 
experimentally established temperature-doping dependent phase diagram, then has to be  
derivable  from such a parent state on the basis of very general symmetry breaking processes.

The quantum vacuum  parent state of the cuprate HTSCs is 
generated by chemical synthesis at high temperatures. It describes a solid solution in an out
of thermal equilibrium high entropy state, which arises from incompatible CuO$_4$ square 
planar stereo-chemical complexes Cu$^{\rm II}$O$_4$ with a Cu-O bond-length of 1.93 \AA \, 
and  Cu$^{\rm III}$O$_4$ with a Cu-O bond-length of 1.83 \AA. Both of them are constrained 
to coexist in  the CuO$_2$ layers, sandwiched  between  the layered charge  reservoirs. 
Chemical reactivity between them, controlled by  the covalency of their Cu - O - Cu bonds, stabilises the overall 
crystal structure  kinetically by  zero point fluctuations of the Cu-O bond-length, 
oscillating between 1.93 and 1.83 \AA. This is evidenced experimentally
in the double peak structured PDF  (pair-distribution function) \cite{Zhang-Oyanagi-2009} and 
in the splitting of the Cu - O - Cu bond stretch mode \cite{Reznik-2006}. The fluctuations 
of the bond-length go hand in hand  with double charge fluctuations  on such deformable
plaquettes favouring  textured meso-structures, as seen in STM-IS 
(scanning tunnelling microscope imaging spectroscopy) \cite{Kohsaka-2007,Kohsaka-2008,Lee-2009} 
(see Figs. \ref{fig:Kohsaka_Nature2008Fig4e}).
The spectral distribution of the binding energy of the bosonic bound pairs on those 
textured  plaquettes shows an isotope shift in  $d^2/dV^2$-imaging studies, upon  
replacement of O$^{16}$ by O$^{18}$ \cite{Lee-2006}. This, together with the strong positive 
isotope effect of the pairing energy $k_BT^*(x)$ \cite{RubioTemprano-2000} and its being 
correlated  to the pressure induced Cu - O bond stretch mode frequency \cite{Haefliger-2006} 
are strong indications that pairing in the cuprate HTSCs derive primarily from their intrinsic crystalline meta-stability.

The local dynamical lattice deformations, involving directionally oriented fluctuating 
Cu-O-Cu bonds, are randomly oriented along   the a and b direction. This breaks the rotational as well
as translational local lattice symmetry on an atomic length scale, and thereby evades any onset 
of long range translational symmetry breaking, which could hinder the stabilization of a 
phase correlated super-fluid  state. 

Given these experimentally established features of spontaneous dynamical local lattice 
instabilities in the cuprate HTSCs, we visualize their high temperature quantum vacuum 
parent state as 
one in which bound spin-singlet pairs pop in and out of existence in an underlying Fermi sea 
of itinerant electrons. As we shall see below, the electrons thereby get absorbed  in 
the construction of localized  bosonic pairs with well defined pairing amplitudes 
fluctuating around a non-zero  amplitude  below  $T^*$. It is   
generated by  a local symmetry breaking in which  bound and unbound electron-pairs of that quantum 
vacuum engage in a locally phase locked quantum superposition. 

The electrons having been eaten up in the construction of transient electron pairs display  purely collective phase fluctuation Goldstone modes, which are controlled by the systems aspiration to condense into a macroscopic coherent quantum state, driven  by a mechanism which could be ascribed to  "Quantum Protection" \cite{Laughlin-Pines-2000,Anderson-2000}.

\begin{figure}[h!t]
\begin{minipage}[c]{6.5cm}
\includegraphics[width=6.5cm]{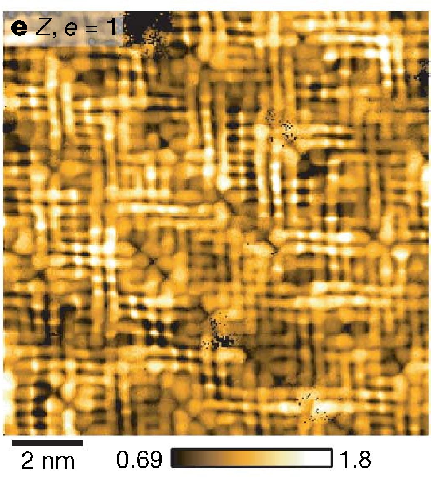}
\caption{Chequerboard segregation of holes (after ref. \cite{Kohsaka-2008}), evidenced in STM 
imaging studies.} 
\label{fig1_Poznan2014}
\end{minipage}
\hfill
\begin{minipage}[c]{6.5cm}
\includegraphics[width=5.5cm]{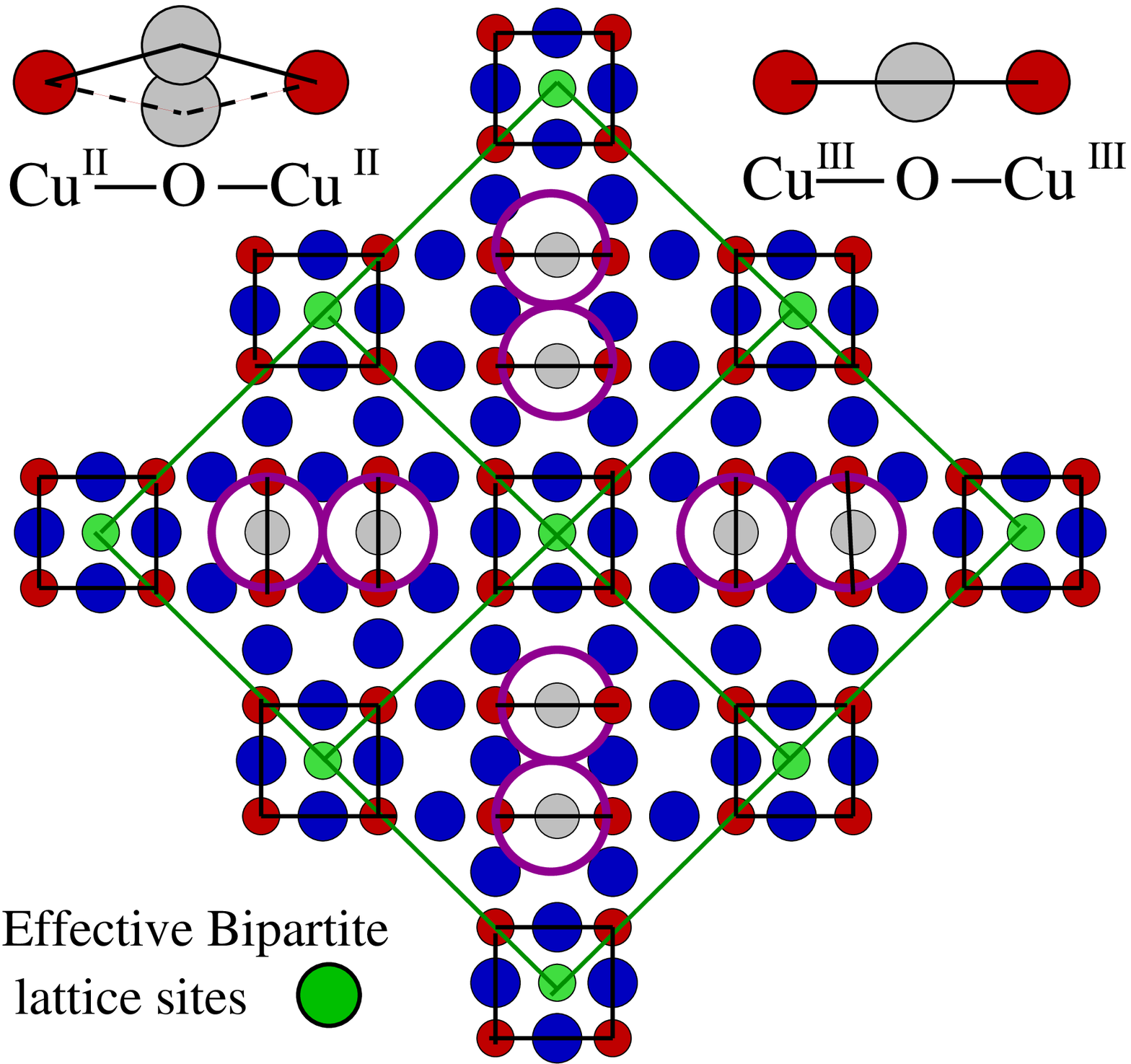}
\caption{An idealized bipartite lattice structure of the texturing of the charge distribution 
of the $CuO_2$ planes. Itinerant electrons move on the sublattice which links effective lattice sites, given by the small filled green circles.}
\label{fig2_Poznan2014}
\end{minipage}
\end{figure}

In order to formulate the scenario of metastability driven pairing of the HTSCs, let us picture 
the spontaneously induced texturing of the CuO$_2$ layers on the basis of an 
idealized regular chequer-board structure bipartite lattice (see Fig.
\ref{fig2_Poznan2014}), composed of two inter-penetrating sub-lattices. 
Itinerant electrons,  experiencing the locally fluctuating covalent Cu - O - Cu bonds in 
the CuO$_2$ layers, get momentarily self-trapped in form of 
resonating Pauling covalent bonds \cite{Pauling-1960}, on dynamically fluctuating molecular 
clusters - square plaquettes composed of  pairs of Cu - O - Cu bonds which fluctuate 
between $Cu^{\rm III}$ - O - Cu$^{\rm III}$ and  Cu$^{\rm II}$ - O - Cu$^{\rm II}$ 
stereo-chemical configurations. We know from 
the theory of  Many Body small polaron physics that a homogeneous state for such a  situation 
is unstable against texturing, driven by local quantum superposition of bound polarons 
(Bipolarons) and quasi-free electron pairs on small clusters, composed of diatomic
molecules \cite{deMello-Ranninger-1997/98}. In the cuprates this manifests itself in form of 
plaquettes composed of locally fluctuating Cu - O - Cu bonds, forming  a sub-lattice "A", which is
embedded in a sub-lattice "B". The effective sites  on that latter are given by plaquettes
composed of rigid  Cu$^{\rm II}$ - O - Cu$^{\rm II}$ bonds  along which we envisage the 
electrons to move as quasi free  itinerant 
particles. In the absence of any such texturing the electrons would form a half filled band 
metal for undoped cuprates (we neglect here any Hubbard on-site repulsion on 
the Cu ions). But in textured structures the freely moving  electrons on sub-lattice 
"B", get momentarily self-trapped when they hop onto  the dynamically deformable plaquettes
on sub-lattice "A".  When considering the dynamically undeformable plaquettes, forming the
effective lattice sites of the sublattice  together with the deformable  cluster on sublattice 
"A", which they surround, two electrons on such effective sites of such a rescaled 
lattice structure exist simultaneously as  bound and un-bound pairs. Driven by the intrinsic
local molecular fluctuations, the exchange between the two is described by a Feshbach
pair resonance coupling \cite{Feshbach-1958}:
$g\sum _{i}\left(\rho^{+}_{i}\tau^-_{i} + \rho^{-}_{i}\tau^+_{i}\right)$. It accounts for the 
transfer of unbound electron 
pairs $\tau^+_{i}=c^{\dagger}_{i\uparrow}c^{\dagger}_{i\downarrow}$
from the metallic substructure on sub-lattice "B" to the 
insulating substructure  on sub-lattice "A", on which these same electrons  get momentarily 
bound into localized hard-core 
bosons in form of localized bound spin-singlet  pairs  $\rho^+_i$ (represented by a 
pseudo-spin $\frac{1}{2}$ operators  $[\rho_i^+,\rho_i^-,\rho_i^z=\rho_i^+\rho_i^- - 1/2]$). 
And vice versa. It is in this dynamical process of transient pairing, induced by spontaneous 
local lattice instabilities that a charge deficiency 
occurring on the  plaquettes of  sub-lattice A, evidences in the STM-IS \cite{Kohsaka-2008} the transiently bound hole  pairs. 

The  interplay between the electrons in itinerant single-particle states and in localized 
two-particle bound states, has been cast into an effective phenomenological Boson-Fermion model 

\vspace {-0.5cm}

\begin{eqnarray}
H_{BFM} =  (\varepsilon_0-\mu)\sum _{i\sigma }
c_{i\sigma }^{\dagger}c^{\phantom{\dagger}}_{i\sigma}
 + (\Delta_B - 2 \mu)\sum _{i}(\rho_{i}^{z}+\frac{1}{2})\nonumber\\
- \sum _{i\neq j,\: \sigma } t (c_{i\sigma
}^{\dagger}c_{j\sigma }^{\phantom{\dagger}} +H.c.)+ g\sum _{i}\left(
\rho^{+}_{i}\tau^-_{i} + \rho^{-}_{i}\tau^+_{i}\right)
\label{eq:BFM-Hamiltonian}
\end{eqnarray}

in the  early  1980ties in an attempt to obtain a super-fluid state of intrinsically 
localized  small Bipolarons. 
The idea behind that proposition was to  work with resonating bipolarons which,
albeit being localized quantities, could exist with locally fluctuating pairing 
amplitudes - generated  by a  crystalline meta-stability. Such a scenario introduces a 
dichotomy of the charge carriers, existing  as both: free fermionic particles with an 
on-site energy $\varepsilon_0$ and  bound bosonic pairs
of them with an energy $\Delta_B$. The two manifestations of the charge carriers,  
coexisting in chemical equilibrium with each other, require that the 
total number of the spin-singlet hard-core bosons and of the fermionic itinerant electrons, 
$n_{\rm tot}=n_{F\uparrow} +n_{F\downarrow}+2 n_B$, 
is conserved. It implies a chemical potential $\mu$,  common to both subsystems. 
$n_B$, $n_{F\uparrow,\downarrow}$ denote the occupation numbers of the hard core-bosons and of 
the electrons with up and down spins.  The strength of the  Feshbach resonant exchange 
coupling $g/t$, in units of the electron hopping integral, is the only free  parameter in this 
scenario. The effect of doping is controlled by the chemical potential which predominantly 
acts on the average density of bosonic charge carriers.
For the cuprate HTSCs, in which the bare conduction electrons form a 
half-filled band in the undoped regime (we neglect any Hubbard on-site repulsion), we put $\varepsilon_0 = \Delta_B = 0$. This  implies that 
the exchange coupling between  unbound and bound electron pairs occurs at the Fermi level of the 
bare itinerant electrons and results in  $n_F = n_B = 1$ when $\mu = 0$.

\vspace{-0.5cm}

 \section{III The anti-correlated doping dependence of $T^*(x)$ and $T_c(x)$}

The salient features of high temperature superconductivity in this crystalline meta-stability 
driven scenario nucleate in the atomic limit of this model 
\cite{Domanski-Ranninger-Robin-1998} and are described by the local physics of the parent state. 
Its  Hilbert space consists of eight eigenvectors, made out of four fermionic 
states $|2\rangle, |3\rangle,|6\rangle, |7\rangle$ and four bosonic states, 
$|1\rangle, |4\rangle,|5\rangle,|8\rangle$,
given by
\begin{eqnarray}
|2\rangle~&=& |c^{\dagger}_{\uparrow} \rangle, \qquad \;|3\rangle~= 
|c^{\dagger}_{\downarrow} \rangle\qquad \; \qquad \; \; \;\;\;E_{2,3} =0 \nonumber \\
|6 \rangle~&=&|c^{\dagger}_{\uparrow} \rho^+\rangle, \;\; \;\;|7 \rangle~=
|c^{\dagger}_{\downarrow}\rho^+ \rangle \; \; \; \, \qquad \;\;\:\; \; E_{6,7} = 0 \nonumber \\
|1\rangle~&=& |0\rangle, \qquad \qquad \qquad \qquad \; \qquad \qquad \; \, E_1=0 \nonumber \\
|4\rangle~&=&(1/\sqrt{2})[e^{+i\frac{\phi}{2}}|c^{\dagger}_{\uparrow}c^{\dagger}_{\downarrow}\rangle-
e^{-i\frac{\phi}{2}}|\rho^+\rangle],\; \, E_4=-g \nonumber \\
|5\rangle~&=&(1/\sqrt{2})[e^{+i\frac{\phi}{2}}|c^{\dagger}_{\uparrow}c^{\dagger}_{\downarrow}\rangle+e^{-i\frac{\phi}{2}}|\rho^+ \rangle],\; \, E_5 =+g \nonumber \\
|8 \rangle~&=&|c^{\dagger}_{\uparrow}c^{\dagger}_{\downarrow}\rho^+\rangle,\qquad  \qquad \; \; \; \;   \; \; \qquad \; \qquad \, E_8=0
\label{eq:atomeigenstates}
\end{eqnarray}
The effect of the crystalline metastability on the electronic structure is  controlled by 
the intra-pair phase rigidity which locks together in a quantum superposition of the two-particle states 
$|4\rangle$ and $|5\rangle$  the phases of (i) bound electron-pairs momentarily occupying 
dynamically fluctuating molecular clusters in form of self-trapped bipolarons (favouring 
an insulating state) and (ii)  unbound  pairs of  delocalized electrons passing momentarily 
through such fluctuating molecular clusters (favouring a metallic state).
At temperatures above a certain $T^* \simeq g$, the thermal fluctuations of the molecular 
clusters destroy this phase locking. Itinerant electrons then are scattered off from
localized  bosonic  bound pairs, which spontaneously appear and disappear on such effective  
lattice sites and thereby loose any Fermi liquid properties. Decreasing the temperature to below
$T^*$, the intra-pair phase locking  (evidenced in the rapid growth of the correlation 
function $\langle c_{\downarrow}c_{\uparrow}\rho^+ \rangle $~=$(1/2)\tanh(\beta g/2)$), 
linking localized bound pairs and unbound pairs on such individual fluctuating molecular clusters  generates the polarizibility of such metastable system, which manifests itself in form: 

(i) of an increasing number of electrons participating  in transient pairing,
$n_p = \langle c^{\dagger}_{i\uparrow} c^{\dagger}_{i\downarrow}c^{\phantom\dagger}_{i\downarrow} c^{\phantom\dagger}_{i\uparrow}\rangle = (1+cosh(\beta g)/(6 + 2 cosh(\beta g)$  which in the zero temperature  limit tends to $n_p =1/2$, which is  twice its  free particle  value $n_p = 1/4$ above $T^*$  

(ii) of the local pair susceptibility in the frequency-zero limit $\chi(i \omega_n \rightarrow 0) = 
\int_0^{\beta} d \tau 
\langle Tc_{\downarrow}(\tau)c_{\uparrow}(\tau)c^{\dagger}_{\uparrow}(0)
c^{\dagger}_{\downarrow}(0)|\rangle =
 \frac{1}{g}[sinh(\beta g) / 3+cosh(\beta g)]$,
which approaches $1/g$ when T drops to below $T^*$. 

This feature illustrates how the incipient polarizability of the parent state above $T^*$ 
generates, through dynamical symmetry breaking,  finite amplitude pairs in an ensemble of 
competing stereo-chemical configurations, once forced by chemical synthesis,
to form a regular crystal structure in 
which the electrons exist simultaneously as itinerant and as trapped into pair states. As
a manifestation of that, the spectral properties of the 
single-particle excitations given by the local Green's function 
\cite{Domanski-Ranninger-Robin-1998} exhibits a characteristic three-pole structure:
\begin{eqnarray}
\label{eq:G_atom}
G_\mathrm{at}(i\omega_n)&=& -\int_0^{\beta}\mathrm{d}\tau\exp^{i
\omega_n \tau}\langle T[c_{\sigma}(\tau)c^{\dagger}_{\sigma}] \rangle \nonumber \\
=[i \omega_n - \Sigma_{at}(i\omega_n)]^{-1} &=&\displaystyle\frac{Z^\mathrm{F}}{i\omega_n} +
\frac{[1-Z^\mathrm{F}]i\omega_n}{[i\omega_n]^2 -g^2},\\
\label{eq:Sigma_atom} 
\Sigma_{at}(i \omega_n) &=& {(1-Z)g^2i \omega_n \over ([i \omega_n]^2 - Z g^2}. 
\end{eqnarray}

The first term in $G_\mathrm{at}(i\omega_n)$, Eq. \ref{eq:G_atom} derives from unbound electrons momentarily  occupying  such
local dynamically fluctuating molecular 
sites with an energy equal to $\varepsilon_0= 0$ and  having a spectral weight 
$Z^\mathrm{F}$~= $2/(3+\cosh\beta g)$. The second term in $G_\mathrm{at}(i\omega_n)$ 
derives from  locally  bound 
electron pairs in bonding and anti-bonding states $|4\rangle$ and $|5\rangle$. These two contributions  have
the structure of the BCS spectral function, with the BCS gap being replaced by the exchange 
coupling $g$.  From the behaviour of $Z^\mathrm{F}$ we 
notice that the  almost temperature independent spectral weight of the central peak at frequency 
zero above $T^*$ abruptly  decreases upon going to  below $T^*$. Correlated to that, the spectral weight $(1-Z^\mathrm{F})$ of
the bonding and anti-bonding contribution abruptly   increases. This feature
of the local Green's function contains the key to the mechanism resulting in the opening the pseudo-gap at $T^*$ in the single-particle density of states
and to the anomalous temperature behaviour of the transport coefficients in the high 
temperature regime above $T^*$. In order to access these properties one has to 
 incorporate this local physics of such fluctuating molecular clusters into an ensemble of 
such clusters in which the bare itinerant electrons move. Given the local nature of 
the Feshbach exchange coupling, a reliable approach to that has proven to be a  Dynamical Mean 
Field Theory (DMFT) analyses \cite{Ranninger-Romano-2010}. We illustrate in Fig. 3  the evolution with temperature of the spectral function of the electrons at the wave-vector $\varepsilon = 0$, which  corresponds to the  Fermi wave-vector before the Fermi surface was destroyed.

\begin{figure}[h!t]
\begin{minipage}[c]{7.5cm}
\includegraphics[width=7.5cm]{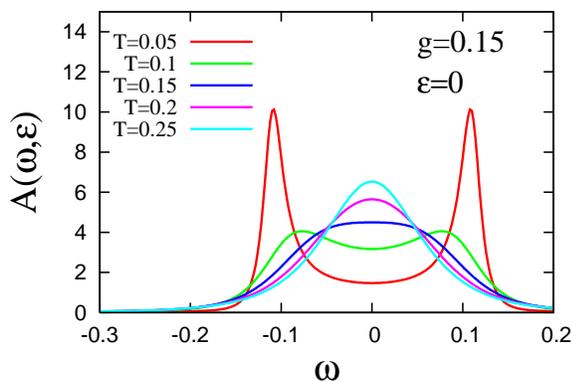}
\end{minipage}
\begin{minipage}[c]{8cm}
\caption{Evolution  with temperature of the spectral function for electrons at the hidden
Fermi surface (after ref.\cite{Ranninger-Romano-2010}), (characterized by an energy 
$ \varepsilon = 0$) for $g/t = 0.15$}
\end{minipage}
\label{fig3_Poznan2014}
\end{figure}

The salient feature of this meta-stability driven pairing and its resulting from that low temperature superconducting, respectively insulating, state is the competition between (i) the 
local intra-pair phase correlations, which  link the bound and unbound pairs on a given site 
(as described by the atomic limit of $H_{BFM}$) and (ii) the inter-pair phase correlation, 
monitored by the electron hopping which links the phases of the bound electron 
pair components of these transient local composite bosonic entities on neighbouring sites. In order
to illustrate that,
let us consider this problem on hand of a  closed ring like structure, which presents both 
of the sublattices in terms of effective composite sites, such as shown in Fig. 4.
\begin{figure}[h!t]
\label{fig4_Poznan2014}
\begin{minipage}[c]{4.5cm}
\includegraphics[width=4.5cm]{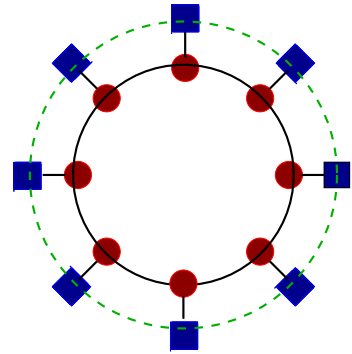}
\end{minipage}
\hfill
\begin{minipage}[c]{8cm}
\caption{A 1D exemplification of the BFM scenario on a closed 8-site ring with composite 
effective lattice  sites. Itinerant electrons sitting on the circular sites in red on sublattice B move on 
the ring via  inter-site hopping and  also hop in pairs onto the pairing centres given by 
the square sites in blue on sublattice B, where they get momentarily trapped in form of localized hard-core bosonic bound pairs.}
\end{minipage}
\end{figure}
Solving this BFM by an exact diagonalization study for such a ring like structure 
\cite{Cuoco-Noce-Ranninger-Romano-2003}, we show on the left panel of Fig. 5 the variation of 
$T^*$ and $T_{\phi}$, which characterizes the onset of local pairing (determining the onset of 
the pseudo-gap state) and of spatial phase correlations (determining the onset of 
superconductivity) for a fixed $g/t=0.5$ as a function of hole doping: $n_B = 4/8 = 0.5$ 
presenting the case of undoped  cuprates and $n_B$ between 3/8  and 0 for the hole doped ones. 
The hole doping $x= 0.5 - n_B$ is tracked by fixing the average  number of bosonic bound pairs $n_B$ on this ring. 
The hole doping appearing symmetric with respect to electron doping with $0.5 \leq n_B \leq 1$ 
in this illustration of $T^*$ and $T_{\phi}$, results from having assumed  
that the pair exchange coupling 
between  bound and  unbound hole pairs is identical to that of bound  and unbound electron
pairs. This is evidently not so, since for electron  doped systems the two   
stereo-chemical configurations which compete with each other are Cu$^{\rm II}$O$_4$ and 
Cu$^{\rm I}$O$_4$, the  former
having a square planar and the latter linear dumbbell stereo-chemical configurations. For  
the sake of the present illustration, where we concentrate on the robust qualitative features of 
the cuprates, we shall  ignore this quantitative  effect on the particle-hole asymmetry 
of the Feshbach exchange coupling.

$T^*$ in this finite size features of the BFM  is determined by an abrupt decrease of the
local intra-pair 
correlations given by $\langle|\rho^+_i\tau^-_i|\rangle$ which shows upon decreasing $n_B$
from 0.5 to 0 (increasing the hole doping from the underdoped to the overdoped systems) a steady monotonously decreasing behaviour. $T_{\phi}$ is determined by an  equally
abrupt increase of the long range phase coherence of the hard core bosonic  bound pairs, 
given by  $\langle|\rho^+_{\bf q} \rho^-_{\bf q}|\rangle$ for $q=0$. 

In Fig. 5 left panel we illustrate how upon decreasing $n_B$ 
from 0.5 to 1/8 (increasing the hole doping from zero up to the optimal doping rate) $T_{\phi}$ increases monotonously until it hits the at the same time monotoneously decreasing $T^*$ around $n_B=1/8$. From there on, upon further increasing the hole doping (decreasing $n_B$), $T_{\phi}$  becomes delimited by the pairing energy $k_B T^*$,  which displays features which are characteristic for  BCS superconductors. 
The minimum of $T_{\phi}$ at $n_B= 0.5$  observed in this  finite size system hints the 
system's  tendency to transit into an insulating Bose glass state composed of spatially 
phase uncorrelated localized transient electron-pairs. For larger values of $g/t$ we find
a more pronounced effect for such an incipient transition, the definite existence of 
which has been  confirmed   by  our 
functional integral formulation of the BFM scenario \cite{Cuoco-Ranninger-2004}. In that 
study, the  onset of the superconducting state for undoped systems ($n_B= 0.5$) happens 
when  $g/t$ is decrease to below $\simeq 2$. For a hole doping corresponding to $n_B \simeq 0.4$, the supercoducting 
state is stabilized for $g/t \leq 1$, which is quantitatively close to the result obtained 
for the finite sized ring-like structure. It is the relatively large  intra-pair phase 
correlations (corresponding to the large value of $T^*$)  which kill in this underdoped regime  $[3/8,4/8]$ for $n_B$ the inter-pair phase correlations. 

In Fig. 5, right panel we illustrate the  variation of  $T^*$ and $T_{\phi}$
as a function of the Feshbach resonance coupling strength $g/t$ for zero hole doping  
($n_B = 0.5$) - representing the undoped cuprates. 

Without having had to assume  any specific hole doping dependent $g/t$ in this BFM capturing 
the  crystalline metastability of the cuprates, the results shown in Fig. 5 
describe qualitatively correctly the physical features observed in the cuprates, which are 
inherent in their temperature - doping dependent phase diagram.  It  accounts 
for the hole doping induced phase transition between an insulating Bose glass  and 
the superconducting state. Approaching the optimal hole doping rate, between $n_B=0$ and 
$n_B = 1/8$, $T_c$  becomes, as we can clearly see from the left panel of Fig. 5 to be 
determined by $T^*$. Upon further increasing the hole doping (decreasing $n_B$ to below 1/8)  the decreasing with  increasing hole doping $T^*(x)$ forces $T_c(x)$ to follow suit as expected for BCS superconductors controlled by the amplitude fluctuations encoded in the doping dependent $T^*(x)$.

\vspace{-0.4cm}

\begin{figure}[h!t]
\begin{minipage}[b]{4.2cm}
\includegraphics*[width=4.2cm]{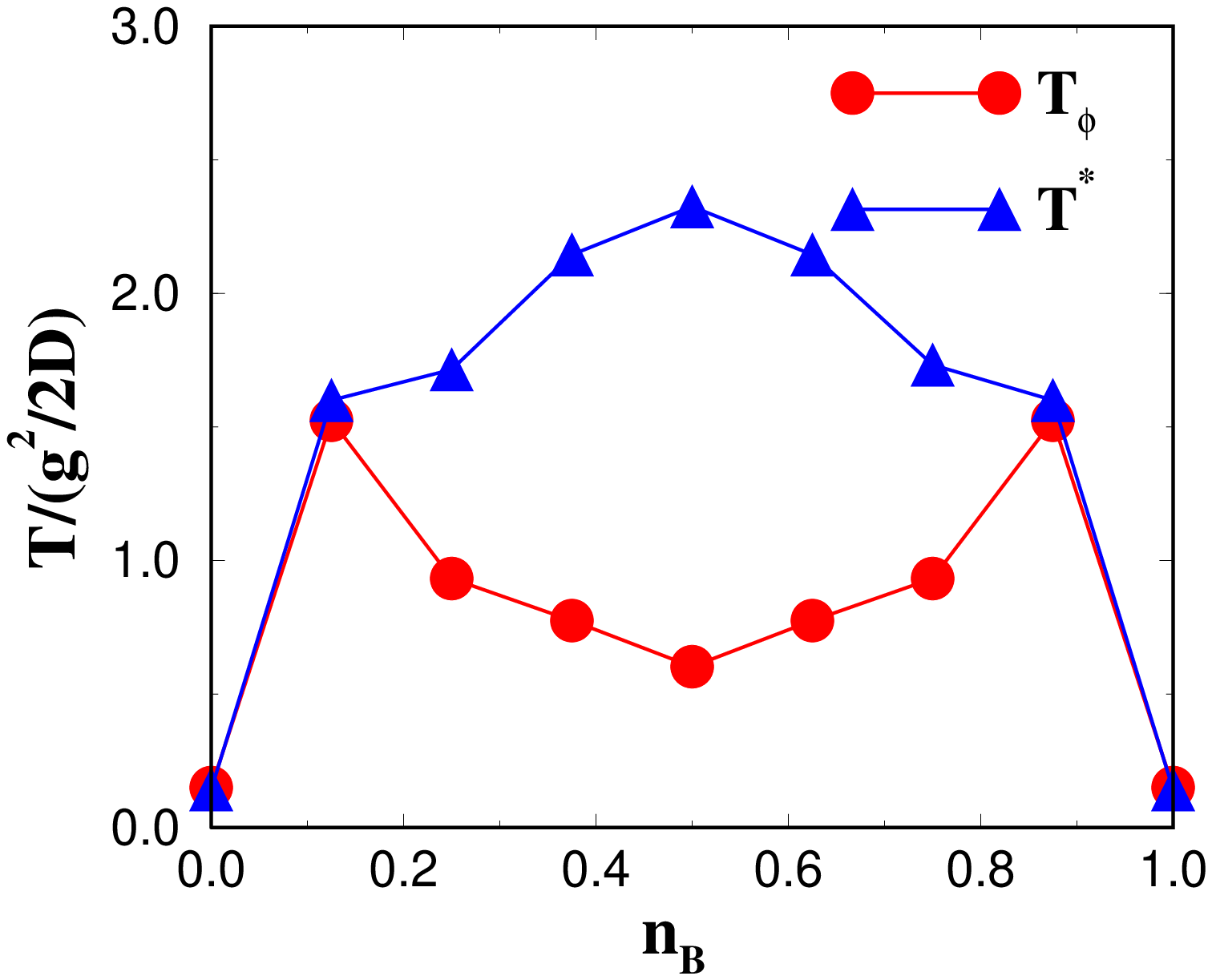}
\end{minipage}
\hfill
\begin{minipage}[b]{4.2cm}
\includegraphics*[width=4.2cm]{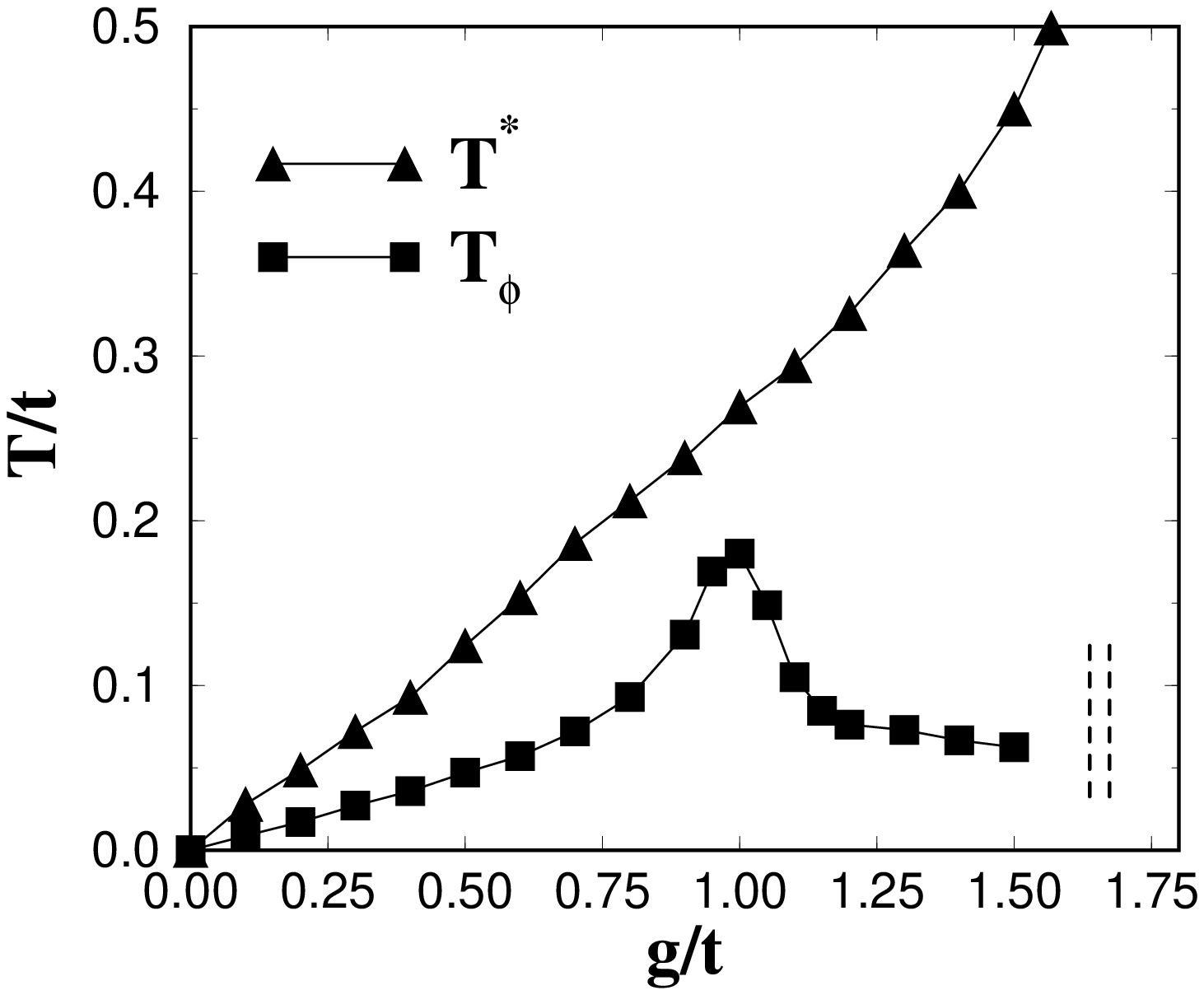}
\end{minipage}
\caption{$T^*$ and $T_{\phi}$ for a closed 8 site ring (after ref. \cite{Cuoco-Noce-Ranninger-Romano-2003}) shown in  Fig. 4 with 
composite effective lattice  sites. The left panel shows the doping dependence  of $T^*$ and $T_{\phi}$ as a function of the average site  occupation by bound pairs for $g/t=0.5$. The right  panel shows the  variation of $T^*$ and $T_{\phi}$ for $n_B = 0.5$ as a function of $g/t$,  which indicates the transition  of the  superconducting phase into the Bose phase glass insulator, when $g/t$ increase beyond unity.}
\end{figure}
\vspace{-1cm }

\section{IV Quantum Protection and the collective excitations of transient localized bound pairs}

The intricate physics of the HTSCs lies in their ability to form a superconducting state,
starting from a system of localized transient spin-singlet electron-pairs with a fluctuating pairing amplitude which manifests itself in the opening of a pseudo-gap in the local single-particle density of states of the electrons. In the process of fabricating such transient  pairs, the initially itinerant electrons inside this pseudo-gap region loose any  quasi-particle features. The spectral weight of electrons, having Fermi-liquid spectral properties, is zero in the energy regime  marking this insulating pseudo-gap state.   With decreasing the temperature, the broad spectral rather structure-less pseudo-gap spectral features of single particle states acquire  upper and  lower Bogoliubov branches \cite{Domanski-Ranninger-2003}. Simultaneously, spatial phase coherence, linking  transient pairs at different sites, generates collective Goldstone phase modes of a super-fluid  condensate of transient real space pairs. 

In order to highlight this intricate interplay between the spectral properties of the 
electrons and of the  resonating transient bound pairs of them, we use a renormalization group procedure, which permits us to decouple the dynamics of the single-particle Fermionic entities  from that of the two-particle  Bosonic entities. 
Using  Wegner's flow equation renormalization group technique 
\cite{Wegner-1994}, we transform  $H_{BFM} = H_0 + H_{int}$ :
\begin{eqnarray}
    H_0 &=& \sum_{{\bf k},\sigma}(\varepsilon_{{\bf k}\sigma} -\mu)c^{\dagger}_{{\bf k}\sigma}
    c_{{\bf k}\sigma}+\sum_{\bf q} (E_{\bf q}-2\mu) b^{\dagger}_{\bf q}b^{\phantom\dagger}_{\bf q}  \\  
    H_{int} &=& \frac{1}{\sqrt N}
    \sum_{\bf k,p}(g_{\bf k,p}b_{{\bf k}+{\bf p}}
    c^{\dagger}_{{\bf k}\downarrow}c^{\dagger}_{{\bf p}\uparrow}+   
    g^*_{{\bf k},{\bf p}}
    b^{\dagger}_{{\bf k}+{\bf p}}c^{\phantom\dagger}_{{\bf p}\uparrow}c^{\phantom\dagger} _{{\bf k}\downarrow}),
\end{eqnarray}
in a sequence of infinitesimal steps, which describes the  flow of the fermionic as well as
bosonic dispersions $\varepsilon_{{\bf k}\sigma}(\ell)$ and $E_{\bf q}(\ell)$, with the  exchange coupling constants $g_{k,p}(\ell)$, getting renormalized down
to zero, at the fixed point where the flow-parameter  $\ell$ reaches infinity. The flow equations which achieve that are given by  $\partial_\ell H(\ell)=[\eta(\ell),H(\ell)]$, 
with  $\eta(\ell)=[H_0(\ell),H(\ell)]$, representing an anti-Hermitian generator and  which has the quality that  $\partial \ell Tr[H(\ell)-H_0(\ell)]^2\leq0$. It is this which assures the total decoupling of bosonic and fermionic fields in the Hamiltonian. The residual interaction between them re-appears in form of the renormalized $\varepsilon^*_{{\bf k}\sigma}(\ell=\infty)$ and $E^*_{\bf q}(\ell=\infty)$, as well as in the renormalized Fermion and Boson operators, given  by 
\begin{eqnarray}
{c^{\dagger}_{-{\bf k},-\sigma}(\ell) \brack c_{{\bf k},\sigma}(\ell)} &=& 
u^F_{\bf k}(\ell) {c^{\dagger}_{-{\bf k},-\sigma} \brack c_{{\bf k},\sigma}} \nonumber \\
&&\mp \frac{1}{\sqrt N}\sum_{\bf q}v^F_{{\bf k},{\bf q}}(\ell){b^{\dagger}_{\bf q} 
c_{{\bf q+k}, \sigma} \brack  b_{\bf q} c^{\dagger}_{{\bf q-k}, -\sigma}}, \\
b_{\bf q}(\ell)&=& u^B_{\bf q}(\ell)  b_{\bf q} + 
\frac{1}{\sqrt{N}} \sum_{\bf k} v^B_{{\bf q},{\bf k}}(\ell)
c_{{\bf k}\downarrow} c_{{\bf q}-{\bf k}\uparrow},
\label{b_Ansatz}
\end{eqnarray}
They determine the spectral properties of the system, as exemplified in ref. \cite{Ranninger-Domanski-2010}.
$u^{F,B}_k(\ell)$ designate the spectral weights of the components of those fermionic, respectively bosonic excitations with well defined individual  particle features. $v^{F,B}_{k,q}(\ell)$, on the contrary, designate the incoherent contributions of their spectral properties.  
The initial fermionic operators $c^{\dagger}_{{\bf k}'\sigma}$, describing 
itinerant electrons with a spectral weight $u^F_{\bf k}(\ell=0)=1$ get renormalized by the appearance 
of an extra term describing a hole travelling together with a bound electron-pair 
$b^{\dagger}_{\bf q}$. 
As the temperature is lowered to below $T^*$, the spectral weight $u^F_{\bf k}(\ell=\infty)$, describing the initially itinerant electrons, tends to zero and the pseudo-gap opens up in their density of states. The missing itinerant fermions have been eaten  up in the process  constructing  spectrally well defined bosonic bound pairs of them, with a dispersion, given by $E^*_{\bf q}(\ell = \infty)$.  The electrons inside the pseudo-gap thereby loose any Fermi liquid  features, as illustrated in Fig. 3. The localized bosonic excitations, characterized by their initial spectral weights $u^B_{\bf q}(\ell=0) = 1, v^B_{\bf q}(\ell=0) = 0$, on the contrary, acquire a well defined collective linear in ${\bf q}$ Goldstone  mode spectrum. It is triggered  by the system's aspiration to condense into a super-fluid macroscopic coherent quantum state, controlled by what has been  coined Quantum Protection \cite{Laughlin-Pines-2000,Anderson-2000}, as the  temperature approaches $T_c$. (see Fig. 6 
right panel).
 
The transition from the superconducting state into the insulating pseudo-gap state is characterized  by the chemical potential  
$\mu^*(\ell = \infty)$ moving out of the renormalized fermionic band $\varepsilon^*_{\bf k}$. It indicates that no fermionic excitations with well defined individual free particle features are left over, as illustrated in Fig. 6 left panel. Simultaneously the transient  bound pairs acquire a free particle behaviour with a $q^2$ spectrum. A more refined inspection of the insulator - superconductor phase transition \cite{Stauber-Ranninger-2007} indicates that at low temperatures it is a first order phase transition, driven by a phase separation involving different  relative concentrations of bound and unbound electron pairs.   In order to determine whether the $q^2$ spectrum obtained for 
$E^*_{\bf q}$ describing transient phase uncorrelated bound electron pairs in the pseudo-gap state actually are itinerant or diffusive modes, we shall have to evaluate  the auto-correlation function of the renormalized boson operators, $b_{\bf q}(l = \infty)$ in Eq. 12.

\begin{figure}[h!t]
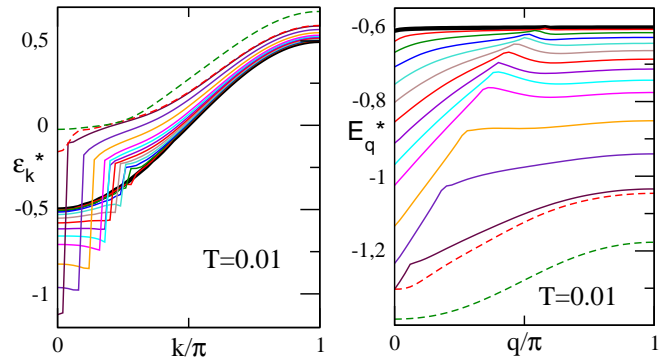

\begin{minipage}[b]{4.2cm}
\includegraphics*[width=4.2cm]{fig6a_Poznan2014.eps}
\label{fig6a_Poznan2014}
\end{minipage}
\hfill
\begin{minipage}[b]{4.2cm}
\includegraphics*[width=4.2cm]{fig6b_Poznan2014.eps}
\label{fig6b_Poznan2014}
\end{minipage}
\caption{The fermionic and bosonic fixed point dispersion $\varepsilon_q^*$  and $E_q^*$  at $T=0.01$ (after ref. \cite{Stauber-Ranninger-2007}) for $g=0.05$ (solid bold line), $g=0.1, 0.15, 0.2, 0.25, 0.3, 0.35, 0.4, 0.45, 0.5, 0.6, 0.7, 0.79$  (solid line) and $g=0.8, 0.9$ (dashed line). The bare values are $\epsilon_{k}(\ell = 0) = -2t cos k$, $E_q(\ell = 0) = -0.6$ for $n_{F\sigma}= n_B = 0.25$ implying $n_{tot}=1$}
\end{figure}
\begin{figure}[ht]
\begin{minipage}[c]{6.25cm}
\includegraphics[width=6.25cm]{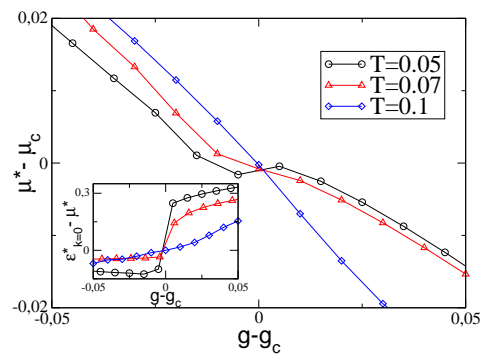}
\label{fig7_Poznan2014}
\end{minipage}
\caption{Variation (after ref. \cite{Stauber-Ranninger-2007}) of the fixed point chemical potential $\mu^*$ with the Feshbach pair exchange coupling $g$ near its critical value
$\mu^*_{c}$ and $g_c$ for a set of different temperatures. The changes from its 
monotonously decreasing behaviour into a non-monotonous behaviour as $g$ varies from  below 
to above a critical $g_c$ indicates the onset of a phase separation driven superconducting 
to non-superconducting state, similar to that of $^3$He - $^4$He mixtures.} 
\end{figure}

\end{document}